\documentclass[conference]{IEEEtran}
\IEEEoverridecommandlockouts
\usepackage{cite}
\usepackage{amsmath,amssymb,amsfonts}
\usepackage{algorithmic}
\usepackage{graphicx}
\usepackage{textcomp}
\usepackage{xcolor}
\def\BibTeX{{\rm B\kern-.05em{\sc i\kern-.025em b}\kern-.08em
    T\kern-.1667em\lower.7ex\hbox{E}\kern-.125emX}}
\begin{document}

\title{A Study of Performance Programming of CPU, GPU accelerated Computers and SIMD Architecture
}

\author{
\IEEEauthorblockN{Xinyao Yi}
\IEEEauthorblockA{\textit{Department of Computer Science}\\
\textit{University of North Carolina at Charlotte}\\
Charlotte, North Carolina, USA \\
xyi2@uncc.edu}
}

\maketitle

\begin{abstract}
Parallel computing is a standard approach to achieving high-performance computing (HPC). Three commonly used methods to implement parallel computing include: 1) applying multithreading technology on single-core or multi-core CPUs; 2) incorporating powerful parallel computing devices such as GPUs, FPGAs, and other accelerators; and 3) utilizing special parallel architectures like Single Instruction/Multiple Data (SIMD).

Many researchers have made efforts using different parallel technologies, including developing applications, conducting performance analyses, identifying performance bottlenecks, and proposing feasible solutions.
However, balancing and optimizing parallel programs remain challenging due to the complexity of parallel algorithms and hardware architectures. Issues such as data transfer between hosts and devices in heterogeneous systems continue to be bottlenecks that limit performance.

This work summarizes a vast amount of information on various parallel programming techniques, aiming to present the current state and future development trends of parallel programming, performance issues, and solutions. It seeks to give readers an overall picture and provide background knowledge to support subsequent research.
\end{abstract}


\section{Introduction} 
\label{sec:introduction}
Over the past decade, High-Performance Computing (HPC) has developed rapidly, aiming to utilize the aggregate computing power of many processing units to solve complex problems. One of the most prominent HPC solutions is parallel computing.

There are various ways to achieve parallel computing. The first method is to take advantage of CPU multithreading. Instead of loading a massive workflow onto a single thread in one core, multithreading technology divides the work into multiple tasks that can be executed by different threads on a CPU, thereby improving execution efficiency. A thread is the smallest unit of CPU scheduling and allocation. Due to hyper-threading technology, the number of threads that can be parallelized is usually twice the number of physical cores.

The second approach involves special hardware designed for parallel computing, such as Graphics Processing Units (GPUs), Many Integrated Core (MIC) architectures, Field-Programmable Gate Arrays (FPGAs), and others. The most notable among these devices is the GPU, so we focus on GPUs as a representative example. The many-core architecture of GPUs excels at data-parallel computation, offering high performance and energy efficiency due to their massive parallel-processing capabilities. The CUDA programming model is used for parallel programming on NVIDIA GPUs. It is widespread but also known to be challenging.

The third method involves unique parallel structures like Single Instruction/Multiple Data (SIMD). SIMD, or vector architectures, have made significant progress over the past decade and are now present in various devices, from commodity CPUs to high-performance computing cores. SIMD is suitable for parallelizing loops and is an efficient solution to the stagnation of Moore's Law.

The parallel technologies used in the three topics discussed above can be classified into three levels. The first level is automatic parallelization or using pre-defined libraries. This approach usually has the least programming difficulty, but optimization is largely hidden from the user. Since this method is designed for high universality, it typically cannot achieve the best performance for a specific application. The second level is explicit parallelism, which uses directive-based programming models like OpenMP and OpenACC to parallelize specific parts of the program. Compared with automatic parallelization, it allows for different optimization strategies for different applications. The third level is manual parallelization, which offers higher flexibility for programmers by using unique programming models or intrinsics to perform more direct operations on the hardware. This method can make better use of the hardware and has the most potential to achieve the best performance.
All parallel techniques are summarized in Table~\ref{tab:table1}.
\begin{table*}
\centering
\begin{tabular}{|l|l|l|l|} 
\hline
& \textbf{low-level} & \textbf{high-level}                                                       & \textbf{library based parallization or auto-vectorization}                                                     \\ 
\hline 
\textbf{multi-core}     & pthread            & \begin{tabular}[c]{@{}l@{}}OpenMP \\ parallel/for directives\end{tabular} &
\begin{tabular}[c]{@{}l@{}}Compiler-based auto-parallization \\ Libraries like MKL\end{tabular}
  \\ 
\hline
\textbf{GPU}            & CUDA, OpenCL            & \begin{tabular}[c]{@{}l@{}}OpenMP target directive\\ OpenACC\end{tabular} & \begin{tabular}[c]{@{}l@{}}Compiler-based auto-parallization \\ Libraries like cuBLAS, cuDNN\end{tabular}
\\
\hline
\textbf{SIMD}           & intrinsics         & \begin{tabular}[c]{@{}l@{}}OpenMP \\ simd directive\end{tabular}          & \begin{tabular}[c]{@{}l@{}}Compiler auto optimization options like -O3 in GCC\\ Some special-designed compilers\end{tabular}  \\                                                  
\hline
\end{tabular}
\caption{Summary of the Parallization Techniques}
\label{tab:table1}
\end{table*}

In order to cover the massive amount of information in these fields and provide a comprehensive overview, we include detailed surveys and summaries based on the related work that is representative in the fields of research. 
This survey provides readers with information on existing challenges and solutions for the performance optimization of these three topics and discusses the potential.
It should be noted that the second topic, parallel optimization with GPUs, occupies the vast majority of this survey. Because parallelization using GPUs is currently the most popular solution.
It's the method I'm most interested in, and my proposal also contains a lot of GPU related stuff.

In the rest of the survey, the architectures are described in Section~\ref{sec:architectures}. Then the multithreading programming in multi-core CPUs is presented in  Section~\ref{sec:CPU}.Section~\ref{sec:PerformanceOptimizationofGPU} will discuss the parallelization on GPU and Section~\ref{sec:simd} is for SIMD. In section~\ref{sec:conclusion} will discuss the conclusion and the proposal.

\section{Architectures} 
\label{sec:architectures}
\subsection{CPU Multithreading and Multicore Architecture}

Multithreading is a technology that enables the concurrent execution of multiple threads in software or hardware environments. A thread is a software-level concept representing the smallest unit of execution within an operating system process. By dividing a large task into multiple smaller tasks and assigning them to different threads, we can achieve concurrent execution and improve overall performance.

On a single-core CPU, multithreaded concurrency is actually pseudo-concurrency, where the CPU rapidly switches between multiple tasks using a scheduling strategy. This approach aims to utilize the CPU core more efficiently by keeping it as busy as possible. Memory interactions can slow down program execution, but multithreading can hide some of this latency through thread scheduling. Hyper-threading technology, primarily used in Intel's high-end CPUs, simulates multiple logical units within a single CPU core. It optimizes thread execution by allowing more efficient use of CPU resources when threads are not fully utilizing the core, thereby improving overall efficiency.

In multi-core CPUs, multithreading allows for the actual parallel execution of threads. Even if one core is fully occupied, other cores remain unaffected and can continue handling thread scheduling. When the number of threads is less than the number of CPU cores, only some cores will work concurrently. If the number of threads exceeds the number of cores, threads are scheduled onto the cores according to specific strategies.

If we suppose there are 12 cores in the CPU, and we create 24 threads. Since the CPU supports hyper-threading, two threads can be loaded into each core for execution, allowing all 24 threads to run simultaneously. If we have 48 threads, the first 24 will be executed while the others are queued. Whenever a core completes its current tasks, the next two threads from the queue are loaded onto that core for execution.

The CPU's cache system is essential for ensuring that threads can access data at high speed. There are two types of Level 1 (L1) caches: L1 instruction cache (L1i), which stores instructions, and L1 data cache (L1d), which stores data. Different threads within the same core share the L1 and Level 2 (L2) caches, while different cores communicate through the Level 3 (L3) cache. A single thread can only run on one CPU core at a time.

Different companies adopt various CPU hardware design models based on cost and performance considerations. However, since we are more concerned with thread scheduling and resource allocation at the software level, we will not delve into the specifics of CPU hardware design models here.

\subsection{GPU Manycore Architecture and Memory Systems of a GPU-Accelerated Computer}

The massive parallelism capability of GPUs is realized through their many-core architecture. The latest NVIDIA Ampere A100 GPU contains 108 Streaming Multiprocessors (SMs), each equipped with 64 FP32 CUDA cores, 64 INT32 cores, and four third-generation Tensor Cores. In total, the GPU has 6,912 CUDA cores, enabling massively parallel processing capabilities.

The GPU's many-core architecture implements a thread execution model known as Single Instruction Multiple Threads (SIMT). With SIMT, the GPU executes instructions in lockstep across multiple threads that process different data. The GPU schedules warps of threads—typically 32 threads per warp—onto its many cores; the warp is the basic scheduling unit of kernel execution.

For programming, the NVIDIA CUDA threading model allows for a much larger number of threads, such as millions of SIMT threads, which are organized in a two-level hierarchy: blocks and grids. At each level—that is, threads within a block or blocks within a grid—these can be structured in 1-D, 2-D, or 3-D topologies. 
%Fig. \ref{fig:CudaMem} shows an example of the CUDA threading model. 
Programmers can determine how to divide computational work and data under this SIMT threading model to fully utilize GPU cores.

%\begin{figure}[h]
%\centerline
%{\includegraphics[width=0.5\textwidth]{Figures/cudaMem.png}}
%\caption{NVIDIA CUDA threading model and GPU internal memory. Picture redrew from: https://www.3dgep.com/cuda-memory-model/}
%\label{fig:CudaMem}
%\end{figure}

A GPU-accelerated computer system can be considered to have two main memory systems: the internal memory hierarchy of the GPU itself and the system memory shared among CPUs, GPUs, and other devices.

%Fig. \ref{fig:CudaMem} also illustrates 
The internal memory hierarchy of a typical NVIDIA GPU is more complex than that of a CPU. Within the GPU, registers and shared memory are on-chip memories that offer fast access times. In contrast, local memory, global memory, constant memory, and texture memory reside off-chip.

Significant trade-offs depend on the type of memory used. Off-chip memories can be accessed by any block or thread, providing flexibility but at the cost of higher latency. On the other hand, on-chip memories are limited in scope—registers are private to each thread, and shared memory is accessible only to threads within the same block—but they offer much faster access speeds.

\subsection{Vector Architectures}

Single Instruction Multiple Data (SIMD) architectures typically consist of two main components: a front-end computer and a processor array.
%, as shown in Fig.~\ref{fig:simd_architecture}.
The processor array is a collection of identical, synchronous processing elements, each with a small local memory where distributed data resides during parallel processing. The front-end computer executes the application program in the usual serial manner but issues commands to the processor array to perform SIMD operations in parallel.

%\begin{figure}[h]
%\centerline
%{\includegraphics[width=0.5\textwidth]{Figures/simd.png}}
%\caption{SIMD Architecture Model. Picture redrew from: https://www.tutorialspoint.com/what-is-simd-architecture}
%\label{fig:simd_architecture}
%\end{figure}

Vector architectures have advanced and changed significantly in the last decade, particularly from dedicated vector architecture machines to general-purpose CPUs with vector extensions. In 2011, Intel shipped general-purpose CPUs with the second version of the Advanced Vector Extensions (AVX2), followed by the AVX-512 extensions in 2013~\cite{AVX512}. The AVX2 extensions contain 256-bit lane widths and are very common in consumer CPUs. The AVX-512 extensions bring 512-bit lane widths into a few high-end consumer CPUs and high-performance CPU product lines. Beyond increasing lane width to 512 bits, AVX-512 brings many advanced features, most notably masking, an ideal solution for stridden loads and stores. The Arm architecture is another notable example of extending general-purpose CPUs to support SIMD. In 2016, Arm announced its Scalable Vector Extension (SVE)~\cite{SVEManual} vector architecture for its high-end Arm CPUs to extend the aging NEON SIMD extensions.
The recent Fukagu supercomputer (ranked \#1 in the November 2021 Top500 list and uses Fujitsu's Arm A64FX CPU with 512-bit vector length) represents the adoption of the Arm SVE architecture in production. 
Arm SVE takes a unique approach to support SIMD architecture of different vector lane widths. Rather than creating a new set of extensions for higher lane widths as Intel does, SVE provides a vector length agnostic model, which allows the same code to run on different hardware with varying vector lengths. In other words, vendors can create vectors as long or short as needed, and the same machine code will run on any hardware.

\section{Multithreading Performance Programming on Multicore CPUs}
\label{sec:CPU}
The methods of converting serial parts of an input program into equivalent multithreaded code can be discussed in three categories based on developer workload and complexity. 
The first category is automatic parallelism. The entire parallelization process is transparent to developers. Developers usually only use automatic parallelization tools or specific options of compilers. The second category is directive-based programming languages, for example, using OpenMP directives. Developers need to include dedicated pragmas in the source code to indicate their intent and, where necessary, add synchronization mutexes and communications to partition the data as needed to generate threaded code. The third category is some low-level programming models—for example, POSIX threads (Pthreads) for multithreading.
Methods in this category allow the developer to control multiple different workflows.

In the following, programming models at different levels will be introduced, including some applications that benefit from these programming models, some common performance problems, and some optimization strategies.

\subsection{Auto-parallelization on CPU}
The advantage of automatic parallelization is that the developer does not have to: 1) 
handle the details of finding loops suitable for work-sharing candidates, and 2) perform data flow analysis to verify correct parallel execution.
It can significantly reduce the developer's workload. Moreover, many legacy codes written in the past few decades need to be reused and parallelized, and automatic parallelization is undoubtedly the most convenient and fast choice. 
Common and popular compilers that support parallelization include Par4All~\cite{amini2012par4all,perez2013proceedings}, Cetus~\cite{eigenmann2005languages,rauchwerger2005languages,dave2009cetus,bae2013cetus}, PLUTO~\cite{bondhugula2008practical}, Intel C++ compiler~\cite{tian2002intel}, AutoPar~\cite{arabnejad2018autopar,dever2015autopar,kalender2014autopar}, etc.

Although compiler-based automatic parallelism is a well-studied area, the underutilization of application parallelism still results in performance levels below what can be achieved by skilled expert programmers.
In 2009, Georgios Tournavitis et al. pointed out the weaknesses of traditional compilers and proposed a novel, integrated approach~\cite{tournavitis2009towards}. First, they use profile-driven parallelism detection, which enables more parallelism to be identified. Furthermore, they replaced traditional target-specific and inflexible mapping heuristics with a machine learning-based prediction mechanism, resulting in better mapping decisions. Their method achieves 96\% performance on the hand-tuned OpenMP NAS and SPEC parallel benchmarks on Intel Xeon platforms.
In 2013, Buğra Gedik et al. proposed a resilient automatic parallelization solution to answer the question of how many parallel channels provide the best throughput in automatic parallelization of general-purpose distributed data stream processing applications~\cite{gedik2013elastic}. They combine workload dynamics and resource availability at runtime to achieve good throughput and short settling times and avoid oscillations and overshoots.
In 2016, Sesha Kalyur et al. proposed an automated parallelization methodology named ParaCite based on the Program Dependence Graph model. ParaCite can address parallelism from a holistic perspective rather than only considering loop-dominant programs.
In 2019, S. Prema, Rupesh Nasre et al. conducted a qualitative and quantitative comparison of five popular automated parallelization frameworks (Cetus~\cite{eigenmann2005languages,rauchwerger2005languages,dave2009cetus,bae2013cetus}, Par4all~\cite{amini2012par4all,perez2013proceedings}, Rose~\cite{quinlan2011rose}, ICC~\cite{tian2002intel}, and Pluto~\cite{bondhugula2008practical})~\cite{prema2019study}. These frameworks differ in their parallelization capabilities and mechanisms. The tests for PolyBench and NAS parallel benchmarks show that ICC stands out clearly. Pluto and Par4all significantly improved performance on the PolyBench benchmark. They also highlighted the need for more sophisticated analyses, user-driven parallelization, and meta-auto-parallelizer.

\subsection{Directive-based Programming Model: OpenMP}
OpenMP was proposed by the OpenMP Architecture Review Board in 1997 and is mainly used for multithreaded programming of shared memory parallel systems.
For the early works, OpenMP was still a very new programming technology at that time. Some of them introduced OpenMP APIs and parallel programming environments~\cite{sato2002openmp,park2001parallel,clark1998openmp}, some of them describe methods to evaluate OpenMP performance~\cite{gonzalez2000applying,furlinger2005ompp,mohr2002design,brunst2005performance,caubet2001dynamic,tanaka2000performance,aslot2001specomp,lin2005static,cappello2000mpi}, some of them attempts to use some techniques or proposed extensions to improve the performance~\cite{gonzalez2001complex,klemm2005proposal}, and some of them work on compilers~\cite{min2001portable,balart2004nanos}.

Although earlier studies are of good reference value, we believe that studies in the past decade are more comprehensive in describing new improvements. Therefore we mainly introduced the OpenMP-related work within the past ten years.

\subsubsection{Implementations}
In 2014, Amit Amritkar et al. described accelerating Discrete Element Method (DEM) using OpenMP~\cite{amritkar2014efficient}. DEM is used for simulating dense particulate systems coupled with Computational Fluid Dynamics (CFD). In two representative examples (DEM-CFD and thermal-DEM), using OpenMP is 50-90\% faster than using MPI.
In 2015, Markus Joppich et al. developed an error correction tool for genome sequencing data named PAGANtec based on the novel PAGAN graph structure and used OpenMP to accelerate it~\cite{joppich2015pagantec}. The performance analysis showed that OpenMP tasks achieved better performance than traditional work-sharing.

\subsubsection{Improvement Techniques}
Nested parallel regions are a potential opportunity to optimize the performance, but the OpenMP runtime relies on heavyweight OS-level threads, so OpenMP does not do well in nested parallel regions. Typically, user-level threads (ULTs) are a lightweight alternative.
In 2019, Shintaro Iwasaki et al. proposed BOLT, a practical ULT-based OpenMP runtime system~\cite{8891628}. With deeply optimized data reuse and thread synchronization and coordination, it achieves similar performance to the widely used leading state-of-the-art OpenMP runtime under flat parallelism and outperforms all existing runtimes under nested parallelism.

Energy minimization for parallel applications is another challenge for multi-core computing systems. In 2015, Rishad A. Shafik et al. designed a novel and scalable energy minimization method~\cite{shafik2015adaptive}. It is based on an iterative learning-based control algorithm that dynamically adjusts voltage/frequency scaling and core allocation based on workload predictions, periodically guided by CPU performance counters. Results on the Intel Xeon E5-2630 platform show up to 17\% lower power consumption for a given performance requirement than existing NAS parallel benchmark methods.

\subsection{POSIX Threads Standard API}
The POSIX threads (Pthreads) standard API creates multiple threads for concurrent processes. It can be applied to multi-processor or multi-core systems, where threads can be implemented at the core level to achieve speed improvement. Gains can also be found in uni-processor systems by exploiting the latency of IO or other system functions that may stall the process.
To use Pthreads, \texttt{pthread.h} header file must be included at the beginning of the script to use all the features of the Pthreads library. To execute a c file, \texttt{-pthread} or \texttt{-lpthread} on the command line must be used when compiling the file.

As early as 1999, Chikara Sakamoto et al. worked on developing the thread library Parallel Pthread Library (PPL) that can provide parallelism and portability at the same time~\cite{sakamoto1998design}. PPL allows parallel execution depending on the system and is a general threading environment.

For C or C++ programs that rely on the Pthreads interface for concurrency, locks are required to avoid data races. When certain memory operations can be reordered relative to the lock, it affects the memory fence required by the lock implementation and therefore has a significant impact on performance. Hans-J. Boehm et al. in 2007 showed that common compiler transformations are actually safe in the presence of pthreads, while they pointed out that reordering constraints are not symmetric for locking and unlocking operations~\cite{boehm2007reordering}.

Multiple levels of parallelization can be achieved using a combination of MPI and Pthreads, MPI for distributed computing (inter-node) and Pthreads for multi-core computing (intra-node). In 2013, Juan F.R. Herrera et al. proposed a model for efficiently map branch-and-bound algorithms on distributed computer architectures using the case of multidimensional Lipschitz Global Optimization~\cite{herrera2013hybrid}. Through the dynamic generation of threads, dynamic load balancing is performed in the intra-node space. The results show improved performance compared to OpenMP and MPI versions used in previous work.

In 2021, Cheng Chen et al. proposed an automatic modeling method to build Program Dependency Net(PDNet) from source code that behaves consistent with the PThread program~\cite{chen2021automatic}. 
PDNet based on Colored Petri Net (CPN) is a multi-threaded program model that describes and distinguishes program dependencies. They overcome two challenges common in PDNet and design a recursive function to solve the problem of computing complex program expressions. Experiments on benchmarks including a subset of the software verification competition SV-COMP demonstrate the usefulness, efficiency, and scalability of their approach.

\section{Performance Programming on GPU-Accelerated Computers} 
\label{sec:PerformanceOptimizationofGPU}
For parallelism techniques in GPU-accelerated computers, we still divide them into three categories according to programming complexity. For automatic parallelization, we will first introduce using the existing GPU function libraries in different fields to achieve acceleration. These fields include medical image processing, matrix-related computation, artificial intelligence, deep learning, etc. The libraries used include cuBLAS, cuDNN, etc. In addition, when using these libraries, developers also propose optimized algorithms for these existing methods or new solutions to actual problems encountered. Especially in some specific fields, the existing libraries cannot fully meet the requirements. 
We then introduce some auto-parallelization methods based on compilers. 
For explicit parallelism, we also introduce how developers in different fields use these directive-based programming languages and the performance improvements. We then focus on optimization strategies that can further improve programming speed.
We mainly introduce some common performance problems and corresponding solutions when using CUDA for manual parallelism. Then we will introduce how developers solve these common problems in specific work.
Although OpenCL is also a framework for heterogeneous platforms, due to space limitations, we only introduce CUDA, which is representative enough.

%Finally, we will summarize the common performance bottlenecks and various solutions on GPU-accelerated computers, hoping to provide suggestions for developers of parallel programming.
\subsection{Library-based Parallization}
%{\color{blue} 
%1. Some evaluation of the libraries, like some benchmarks.
%2. Implementation in different areas.
%}
Using the existing GPU function libraries is undoubtedly the fastest and easiest way to realize parallelism. Calling the functions in libraries makes using the GPU to accelerate an existing algorithm whose performance is not satisfactory possible. Users do not need to know the specific parallel strategy and parallel programming methods. The entire parallel process is analyzed and implemented by functions in the library, which is transparent to the users. 
Many companies and laboratories are committed to developing GPU function libraries to provide users with fast parallel solutions. Some of them are listed below:
\begin{itemize}
    \item CUDA-X launched by NVIDIA: It is a library collection covering multiple application fields built on CUDA. It contains math libraries, parallel algorithm libraries, Image and Video Libraries, Communication Libraries, Deep Learning Libraries and some libraries used on partner platforms.
    
    \item MAGMA launched by Dongarra's Group: MAGMA is a dense linear algebra library that can be applied in heterogeneous/hybrid architectures ("Multicore+GPU" systems). 
    
    \item GPULib launched by Tech-X Corporation: GPULib is designed as a mathematical library for Interface Description Language (IDL) and Matrix Lab (MATLAB).
    
    \item OpenVIDIA launched by Eyetap Personal Imaging Lab: OpenVIDIA is designed as a GPU accelerated Computer Vision Library.
\end{itemize}

CUDA-X contains function libraries in many fields, which are improved after a long development period and are commonly used by developers. Therefore, in this paper, we focus on applying CUDA-X in multiple fields.

\subsubsection{Libraries Related to Mathematical Algebra}
In CUDA-X, libraries such as cuBLAS, cuFFT, cuRAND, cuSPARSE, cuTENSOR, cuSOLVER, etc., which are specially used for mathematical computation, belong to Math Libraries set. CuBLAS provides some basic linear algebra subroutines (BLAS). CuFFT is a proprietary Fast Fourier Transforms implementation for Nvidia's Graphics Processing Units. CuRAND includes some random mathematical functions. CuSPARSE has some basic linear algebra for sparse matrices. CuTENSOR includes the functions for linear tensor algebra, and cuSOLVER provides some dense and sparse directive solvers.

%Many researchers have evaluated the performance of these libraries and provided some improvement strategies to make the exist functions faster, and many developers in various areas are also using these libraries and getting better performance.

Some researchers have evaluated the performance of these libraries and provided some improvement strategies to make the existing functions faster.
In 2008, Sergio Barrachina evaluated the kernels for implementing performing large numbers of arithmetic operations, which is known as Level 3 BLAS~\cite{barrachina2008evaluation}.
Different from the previous evaluations by other researchers~\cite{fatahalian2004understanding,larsen2001fast,moravanszky2003dense}, they evaluate the performance of the latest APIs of cuBLAS on the new generation of hardware (GeForce 8800 Ultra). The APIs they tested included sgemm(), which does the single-precision general matrix multiplication, ssyrk(), which multiplies asymmetric single precision matrix by its transpose and adds a second matrix, strsm(), which can be used to solve the triangular linear systems.
After evaluating, they realized that these APIs were not all well improved, so they came up with some better performance alternatives. These schemes include improving the application of padding, applying Level 3 BLAS kernels on top of sgemm(), and using some methods to reasonably split computing tasks between the CPU and GPU.
Their experimental results show that the methods they take all improve the operation speed of the three APIs.
In 2011, Maxim Naumov from NVIDIA showed in a white paper how to use cuSPARSE and cuBLAS to speed up incomplete-LU and Cholesky preconditioned iterative methods. The operations mentioned include sparse matrix-vector multiplication and sparse triangular solve.
Their experiment results show that the sparsity pattern of the coefficient matrix has a significant impact on performance, and overall, using cuSPARSE and cuBLAS on GPU can achieve about 2x faster than using MKL on CPU.

In 2015, Jose Luis Jodra et al. evaluated cuFFT~\cite{jodra2015study}. They analyzed two essential characteristics of cuFFT and two configuration parameters that affect execution efficiency. Then they came up with patterns to help researchers using the cuFFT library determine which parameters to use to get the best performance, including time and memory consumption.
Similar work was done by David Střelák et al. in 2018, who proposed a tool called cuFFTAdvisor, which can analyze given constraints of input size and plan settings and automatically set the optimal configuration of the cuFFT library~\cite{stvrelak2018performance}. Their experiments show that an average 6.9x speedup can be achieved using their tool.
In 2014, Hovhannes Bantikyan used polynomial multiplication to compare several well-known fast big integer arithmetic libraries (.net BigInteger, GMP, IntX) with cuFFT~\cite{bantikyan2014big}. Their experiments showed that cuFFT outperformed all other methods when they dealt with $2^{15}$-bit integers.

%{\color{blue} Here need more reference for evaluation(some benchmarks?)}

These libraries are also widely used in various industries.
In 2010, Bo Zhang et al. demonstrated the use of cuBLAS and cuLA in bioluminescence tomography(BLT)\cite{zhang2010cublas}. CULA supports more advanced linear algebra calculations than CUBLAS, and it is built on top of CUBLAS. Matrix multiplication in BLT yields the most significant speedup in experiments.
In 2012, Beata Bylina et al. applied cuBLAS on GPU and BLAS on CPU, respectively, to WZ factorization~\cite{bylina2012gpu}. WZ factorization is a dense, square, non-structured matrix factorization algorithm. They first rewrote the WZ factorization as an operation on blocks of matrices and vectors, making it possible to use cuBLAS. NVIDIA Tesla C2050 GPUs test results show GPU implementations up to 6x faster than multithreaded CPU solvers for large matrices. They also proposed that using cuBLAS did not compromise the accuracy of the results.
In 2020, N. Ragoomundun et al. applied cuBLAS to radio astronomy~\cite{ragoomundun2020cublas}. They used functions in the cuBLAS library to perform batched multiplication of matrices to compute the full correlations in the frequency domain. Although the cuBLAS-based implementation does not offer significant performance advantages compared to the implementation using xGPU~\cite{clark2013accelerating}, this work represents a solution that can be developed in a short time and implemented quickly.
In 2013, Yong Huang et al. applied the cuSPARSE and cuBLAS libraries to realistic modeling and deformation of soft tissue~\cite{yang2013fem}. They adopt linear FEM (Finite Element Method) and sparse matrices stored in CSR format, using cuSPARSE and cuBLAS to implement the computation process on the GPU. The experimental results show that the accelerated method effectively enables realistic and fast soft tissue modeling and deformation simulation. Their research has led to the development of new clinical applications such as virtual surgical simulators.

\subsubsection{Libraries Related to Deep Learning}
In addition to some libraries related to mathematical computing, libraries related to deep learning are the most popular. Such as NVIDIA cuDNN~\cite{chetlur2014cudnn}, NVIDIA TensorRT~\cite{vanholder2016efficient}, NVIDIA Riva~\cite{NVIDIARiva:web}, NVIDIA DeepStream SDK\cite{DeepStreamSDK:web}, and NVIDIA DALI~\cite{DALI:web}. CuDNN is a GPU-accelerated library of primitives for deep neural networks, with optimized routines for deep learning workloads. Now it is used by most deep learning frameworks for their GPU operations. We investigate related evaluations and applications of cuDNN below to give an overall picture of how the libraries are applied in deep learning.

The developers of cuDNN, Sharan Chetlur et al., published cuDNN in 2014~\cite{chetlur2014cudnn}. They point out that cuDNN can be easily integrated into existing frameworks like TensorFlow, PyTorch, or Caffe2.
Their experiments show that when integrating cuDNN on Caffe, using cuDNN can improve performance by 36\% over the standard model.
In 2019, Marc Jordà et al. conducted a performance evaluation of the convolution algorithm provided by cuDNN~\cite{jorda2019performance}. They evaluate the performance of the convolution algorithm by changing the configuration of the convolutional layers of common CNNs, extracting the input size, filter size, number of filters, and input and filter depths in these configurations. 
Experiments are conducted on NVIDIA V100 GPU, and the experimental results show that filter size and the number of inputs are the most important parameters when choosing a GPU convolution algorithm for 32-bit FP data. For 16-bit FP data, the best performance is obtained when leveraging NVIDIA Tensor Cores.

% Put cuDNN evaluation here:

% Put implementations here:
In 2015, Di Zhao et al. applied cuDNN to MRI~\cite{zhao2015gpu}. MRI is an essential tool for diagnosing and screening for brain carcinoma. Convolutional Neural Networks (CNN) can help doctors improve diagnosis efficiency.
Using cuDNN, they developed a fast CNN that can perform segmentation in minutes with the same accuracy as a traditional CNN that takes hours.
The problem of 3D brain tumor segmentation is solved by segmenting the 2D slice in the axial direction one by one. So the 3D CNN problem is computed by 2D convolution functions in cuDNN.
Hannes Fassold et al. improved YoloV3 C++ to propose OmniTrack with better performance in 2019~\cite{fassold2019omnitrack}. OmniTrack is a deep learning-based object detector commonly used for automatic inspection and tracking of objects, including people, animals, text, and logos in videos. The original YoloV3 C++ implementation can only complete one inference call at a time on a particular GPU because this GPU uses a single CUDA stream and CuDNN handle. OmniTrack changes these global variables to thread-local variables so that each CPU thread has its own private CUDA stream and CUDNN handle. This enables the network instance to create its persistent worker thread, significantly improving efficiency.

%Put the improvement here:
Since cuDNN is a relatively new library, many researchers have made improvements on the basis of cuDNN to obtain better performance or be better applied to their fields.
In 2018, Yosuke Oyama et al. proposed a thin wrapper library for cuDNN, named u-cuDNN~\cite{oyama2018accelerating}.
It can split the mini-batch computation of a layer into multiple micro-batches. Thus, these computing tasks are distributed to single or multiple GPUs. Reducing workspace requirements is how u-cuDNN implements faster algorithms. Their experimental results show that on P100-SXM2 GPU, u-cuDNN can achieve 1.63x speedup for AlexNet and 1.21x for ResNet-18. On the V100-SXM2 GPU, it can achieve up to 4.54 times the DeepBench convolutional layer acceleration.

Due to the limited memory capacity of the GPU, to achieve large-scale CNN computation on the GPU, data allocation must first be performed on the CPU, and the data movement overhead between the CPU and GPU cannot be ignored. So in 2017, Yuki Ito et al. proposed an extension of cuDNN called ooc\_cuDNN~\cite{ito2017ooc_cudnn}. It reduces communication overhead through pipeline processing and division-size optimization based on the performance model. The original cuDNN handles the divided computational tasks. They implemented a CNN that required 60GB of memory for computation on a Tesla P100 GPU with 16GB of memory and only saw a 13\% performance drop compared to in-core cases using the original cuDNN. Moreover, ooc\_cuDNN gets 1.4x faster compared with using unified memory.

Although it is impossible to mention all related works in the limited space, the work mentioned above is enough to show the importance that researchers and developers place on GPU programming libraries. Compared to traditional CPU serial programming, the performance gains achieved with GPUs are considerable.

\subsection{Auto-parallization Based on Compielrs}
There are two common automatic parallelization methods based on compilers. One is to convert serial source code into parallel programs in CUDA, OpenCL, OpenMP, etc. The other one is to use these parallel frameworks and their runtime libraries in the backend, which is not visible to users.

In 2013, Guodong Han et al. introduced Japonica, a compiler framework and runtime system that helps Java applications take advantage of heterogeneous systems~\cite{han2013java}. Japonica can automatically parallelize loops and schedule workloads efficiently. With simple user annotations, sequential Java source code will be analyzed, translated, and compiled into CUDA kernels and multiple Java threads running on GPU and CPU cores, respectively. Their test results show that Japonica runs ten times, 2.5 times, and 2.14 times faster than the single-threaded CPU, GPU, and multi-threaded CPU versions.
In 2018, Wim Vanderbauwhede et al. proposed a compilation method that automatically converts FORTRAN 77 legacy code into OpenCL-accelerated programs through whole-program analysis~\cite{vanderbauwhede2018domain}. They also improve the performance of the generated programs by eliminating redundant data transfers. Test results on the 2D Shallow Water Model (2DSW) and Large Eddy Simulator for Urban Flows (UFLES)  show that the generated  OpenCL programs are nine times and 20 times faster than the original FORTRAN program, respectively, which is comparable to the performance of the manual porting program. 
In 2021, M. Usman Ashraf proposed an adaptive and automatic parallel programming tool (AAP4All) for homogeneous and heterogeneous computing systems~\cite{ashraf2021aap4all}. AAP4All can automatically recognize the computer system architecture and then generate specific parallel code, such as MPI, OpenMP, CUDA, or OpenACC, according to that particular detected system. In addition to better performance, the generated program also reduces power consumption.

\subsection{Using Directive-based Language to Achieve Parallelism}
Directive-based GPU programming is regarded as a simple GPU programming method, allowing the terminal programmer to have more control over a specific code block than directly calling predefined functions mentioned in the previous chapter.
Establishing the directive-based language standard for GPU programming has greatly interested researchers.
OpenMP, proposed in 1997, is considered to be a typical representative of directive-based programming~\cite{dagum1998openmp}. OpenMP is good at handling multi-threaded parallelism on a shared memory computing platform and focuses more on CPU power. In 2011, several compiler vendors Cray, CAPS, Nvidia, and PGI, released the OpenACC standard in order to better provide shortcuts for parallel programming of heterogeneous computing systems(CPU+GPU)~\cite{openacc2011openacc}. It follows almost the same design principle as OpenMP, with some specific directives and clauses to create programs for heterogeneous systems without managing accelerator programming languages. In 2013, the \texttt{target} construct was proposed in OpenMP 4.0, which can move the code executed by the CPU threads to the target device and redistribute the required data~\cite{openmp:web}. This improvement supports the application of OpenMP on the GPU.

\subsubsection{Evaluations and performance improvements of OpenMP offloading}
The design of OpenMP offloading is very flexible, and the advantage is that it can be flexibly applied to various scenarios.
To analyze and optimize the performance of the applications, adequate tools support is essential.
The OpenMP tools interface (OMPT)~\cite{eichenberger2013ompt} was described in a technical report published by the OpenMP Architecture Review Board (ARB) in 2013 as a possible future extension of the OpenMP specification. But the most exciting new feature in OpenMP 4.0- the \texttt{target} construct is not considered. In 2015, Tim Cramer proposed a more informative extension, including detailed definitions of necessary runtime events, the signatures, and the inquiry functions~\cite{cramer2015performance}. They discuss the factors that need to be considered in performance analysis work and show the advantages and disadvantages of OMPT. Their work can help users better understand performance issues and benefit from better code tuning performance measurement metrics.
Similar work was done by Veronica G. Vergara Larrea in 2016, who examined the validity of the OpenMP 4 offload model as a pervasive technique for programming modern node architectures~\cite{larrea2016early}.
Using Cray and the Intel compiler, they tested the Jacobi kernel on OLCF's Titan supercomputer (equipped with an NVIDIA Kepler K20X GPU). When using the GPU, it is about 50 times faster than using only the CPU. They also point out that using the data directive is essential for performance.

In 2015, Carlo Bertolli et al. integrated the complexity of control loops into Clang by limiting its support to OpenMP-related functionality and comprehensively reporting benchmarks and performance analysis results of complex application kernels~\cite{bertolli2015integrating}. They demonstrate optimizations for specific code patterns in Clang's code generation scheme alternative to the control loop, resulting in improved performance.

In 2018, Jose Monsalve Diaz et al. evaluated compilers (including GCC, Clang/LLVM, IBM XL, and Cray CCE compilers) on DOE supercomputers Titan and Summitdev for OpenMP 4.5 target offload support~\cite{diaz2018evaluating}. Their tests evaluated the use of OpenMP and exposed ambiguities in the OpenMP 4.5 specification. Their work helps compiler developers fix the ambiguities and let programmers realize the importance of the proper methodology.

%\subsubsection{Some performance issues and the improvement of OpenMP offloading}

Although OpenMP has shown outstanding performance advantages in many applications and evaluations, it still has many problems. 

As we mentioned above, the offloading of OpenMP is very flexible. Although this design allows it to be applied freely in various scenarios, modern compiler front-ends have limited analysis and transformation capabilities. It will choose the correct but conservative code by default. Thus the actual syntax for OpenMP offload may significantly affect performance. The optimization of OpenMP has also attracted much attention, mainly focusing on optimizing the compiler.
In 2019, Johannes Doerfert et al. proposed the use of semantic source code analysis instead of relying on the user-provided syntax and prototyped their solution in the OpenMP compiler LLVM/Clang~\cite{doerfert2019tregion}. Test results on the Rodinia benchmark show that their design achieved up to 1.55 times speedups.

Another problem with OpenMP offloading is that when dealing with sequential code in an offloaded region, the commonly used approach is to map the sequential region to the CPU while executing the parallel region on the GPU. When there are multiple parallel regions in a program and serial regions isolate these parallel regions, multiple calls to the GPU are required to execute the offloaded code regions, and very complex data movement between the CPU and GPU can also occur.
Another solution is to use multiple blocks for a single kernel. The design concept refers to the dynamic parallelism in CUDA, using the GPU to call the parallel area, and making the sequential code execute sequentially through the if statement. This approach results in a lot of unnecessary thread synchronization.
Carlo Bertolli et al. proposed a new thread coordination scheme in 2014~\cite{bertolli2014coordinating}. 
This scheme can launch a single fully parallel GPU kernel to offload regions of arbitrary code sequences, including sequential and parallel regions. Moreover, it only affects the OpenMP code generation component so that it can be integrated into the LLVM compiler in a modular fashion. They program CUDA blocks and threads without returning execution results to the host. Although their test results did not show a significant performance improvement, their research showed us the possibilities of using this approach and the potential for subsequent performance improvements based on this approach.

In a CPU+GPU heterogeneous system, there may be many issues in the data movement between the CPU and GPU, such as data mapping issues caused by the incorrect usage of target offloading constructs and data synchronization problems between host and device, use of uninitialized memory, etc.
In 2021, Lechen Yu et al. proposed ARBALEST and on-the-fly data mapping issue detector for OpenMP applications~\cite{yu2021arbalest}. They mentioned that the root cause of the data mapping issues is that the \texttt{map} and \texttt{nowait} clauses in the target offloading constructs are not set correctly. For each variable mapped to the accelerator, ARBALEST leverages a state machine to track the visibility of the last write. Their evaluation results show that ARBALEST has higher accuracy than the other four dynamic analysis tools (Valgrind, Archer, AddressSanitizer, MemorySanitizer), and its execution time overhead is comparable to the state-of-the-art dynamic analysis tools.

\subsubsection{OpenACC implementations and evaluations}
OpenACC is an API consisting of compiler directives for offloading loops and regions of C/C++ and Fortran code to the accelerator.
Unlike OpenMP, it is specially designed for CPU+GPU heterogeneous systems.
After being proposed in 2011, it has attracted much attention.
In 2012, JA Herdman et al. applied OpenACC, OpenCL, and CUDA to Lagrangian-Eulerian explicit hydrodynamics mini-application, respectively~\cite{herdman2012accelerating}. On a GPU-based cluster (Cray XK6), they compared the performance, programmer productivity, and portability of these three methods, and then they realized that OpenACC provides comparable performance to un-optimized OpenCL and CUDA but dramatically reduces the programming difficulty. This finding validates the significance of OpenACC being proposed. It is an attractive and viable programming model for future accelerator devices.
John M. Levesque et al. transformed a turbulent combustion solver called the S3D program that initially only applied MPI into a hybrid MPI/OpenMP parallelism and then used OpenACC to generate code for the accelerator~\cite{levesque2012hybridizing}. S3D is the first complete production application ported to OpenACC, showing what can be achieved using the directive-based method.

In 2014, Jiri Kraus et al. used OpenACC to accelerate the flow solver Zonal Flow Solver (ZFS)~\cite{kraus2014accelerating}. ZFS is mainly used for numerical prediction of internal and external, industrial, and biomedical flows and the study of fundamental problems in fluid dynamics, such as Near-field acoustic simulation of a jet engine. They also applied some optimization methods, including using the texture cache and reducing the occupancy rate to improve the hit rate of the texture cache. Experimental results on Tesla K40m show that the speedup can be up to 2.36 times compared to the serial version.

Lattice Boltzmann (LB) is widely used in computational fluid dynamics to describe two and three-dimensional flows.
In 2016, Enrico Calore et al. described a state-of-the-art LB algorithm written entirely in OpenACC, and they also described the accurate performance measurements and evaluations of practical portability improvements~\cite{calore2016performance}. Regarding performance, their test results on Tesla K40 and K80 GPUs show that the overall performance may be 50\% of the performance achieved by CUDA.
In 2017, Ao Xu et al. also applied OpenACC to LB simulations of fluid flow, heat, and mass transfer and obtained orders of magnitude speedup compared to serial algorithms~\cite{xu2017accelerated}. They applied optimizations such as optimizing data layout, minimizing memory access frequency, and adjusting the number of gangs and vector lengths.

In 2019, Haowei Zhang et al. optimized the three-dimensional Tokamak magnetohydrodynamical code (CLT) using OpenACC and MPI techniques~\cite{zhang2019acceleration}. They tested OpenACC's performance on a single NVIDIA TITAN Xp and TITAN V. With TITAN Xp, the running time was nearly 256 times faster compared to Intel® Xeon® Gold 6148F CPU. The TITAN V is even 54\% faster than the TITAN Xp.
In 2021, Jae Youp Kim et al. applied OpenACC to the time-consuming microphysics scheme Weather Research, and Forecasting (WRF) single-moment 6-class microphysics scheme (WSM6)~\cite{kim2021gpu}. They adopted optimizations such as minimizing the data transfer between the CPU and GPU, and reducing the waste of GPU threads during computation. Experimental results on Tesla V100 show that WSM6 achieves 5.71x speedup when porting the entire model to GPU, except for I/O communication.

As we mentioned above, compiler optimization is the key to optimizing directive-based programming languages.
In 2012, Ruymán Reyes et al. developed an OpenACC compiler, accULL, which is an implementation based on a combination of source-to-source compilers and a runtime library named Frangollo, providing support for both OpenCL and CUDA platforms~\cite{reyes2012accull}. AccULL is a programming language-level compiler, making it easier for developers to implement algorithmic enhancements instead of spending much time on device-specific code.
Then, in 2013, their research team evaluated the only two commercial OpenACC implementations available at the time (PGI and CAPS) and accULL~\cite{reyes2013preliminary}.
Across the performance tests for three test cases (A LU decomposition (LUD), a thermal simulation tool (HS), and a dynamic programming algorithm (PF)) on NVIDIA C2050, accULL outperformed PGI and CAPS. Although overall, their tests for OpenACC show worse performance than CUDA 4.1 implementation, in one of the test cases, HS achieves 67\% of CUDA's implementation on accULL, which is a satisfactory result.

%{\color{blue}evaluation}
Evaluating the performance, evaluating the correctness of the OpenACC implementation, and determining whether it conforms to the specification are also critical.
In 2014, Cheng Wang et al. designed an OpenACC compiler validation test suite and infrastructure that is used to validate and verify different OpenACC compiler implementations for conformance, correctness, and completeness for OpenACC 1.0~\cite{wang2014validation}. 
Their design has a test infrastructure in addition to some test code. The main contribution is that their test suite helps compiler developers to validate their compiler’s conformance to the specification. In OpenACC 2.0, most of the ambiguities they reported were resolved.
In the same year, Amit Sabne et al. evaluated the impact of OpenACC various compiler optimizations and OpenACC program settings on NVIDIA CUDA, AMD GCN, and Intel MIC architectures. Their analysis of 12 programs highlights the role of specific optimizations for individual architectural features and provides a performance portability matrix showing the potential benefits of highly optimized applications.

\subsubsection{Performance comparison or translators}
Although OpenMP and OpenACC share the same goal of allowing users to specify parallelism through directives, code porting is considered challenging due to the different parallelisms and memory hierarchies on different architectures. So many researchers make efforts to translate between OpenACC and OpenMP.

In 2016, Guido Juckeland et al. converted the OpenACC Suite (consisting of 15 applications written in C or Fortran) in the SPEC ACCEL Benchmark into OpenMP 4.5~\cite{juckeland2016describing}. Compared with OpenACC, OpenMP can be applied to more acceleration platforms. Their goal is to report the difficulties of the porting process and provide guidance on how to translate between them or implement parallelism with them. Although they could not provide performance results, their research can help the community and compiler vendors understand how users plan to write OpenMP 4.5 applications in a performance-portable style.

\subsection{Using CUDA to Achieve High Performance for NVIDIA GPUs}
\subsubsection{Implementations in different areas}
\label{subsec:Implementationsindifferentareas}
In June 2007, NVIDIA launched CUDA (Compute Unified Device Architecture), which aims to use the advantages of CPU and GPU fully.
Although the directive-based programming languages OpenMP and OpenACC can significantly simplify programming complexity, CUDA has an absolute advantage in performance compared with them. This is because CUDA is more flexible; e.g., it provides fine-grained synchronization primitives such as thread synchronization and atomic operations.

In 2008, one year after the release of CUDA, Michael Garland et al. concluded the work using CUDA to parallelize and accelerate computation in various domains from many researchers~\cite{garland2008parallel},
including Molecular Dynamics~\cite{anderson2008general}, Numerical Linear Algebra~\cite{volkov2008lu}, Medical Imaging~\cite{stone2007gpus}, Fluid Dynamics~\cite{phillips2008multi}, Seismic Imaging. Their experimental results all achieve significantly better performance on the GPU than the CPU implementation. For example, Highly Optimized Object-Oriented Molecular Dynamics (HOOMD) implemented in CUDA and executed on a single GeForce 8800 CTX is equivalent to using 32 CPU cores.
In 2009, Julien C Thibault et al. described a Navier-Stokes solver for incompressible fluid flow using a multi-GPU-equipped desktop platform~\cite{thibault2009cuda}. Projection algorithms for solving incompressible fluid flow equations are divided into different CUDA kernels. Using a Quad-GPU platform, CUDA can achieve two orders of magnitude speedup compared to a serial CPU implementation.
In 2011, Martin Burtscher et al. used CUDA to implement the Barnes Hut force-calculation algorithm, which is widely used in n-body simulations such as modeling the motion of galaxies~\cite{burtscher2011efficient}. It repeatedly builds and traverses irregular tree-based data structures, performs many pointer-chasing memory operations, and is typically expressed recursively. The primary meaning of this work is to show that GPUs can be used to accelerate irregular code, not just regular code.
In 2012, S. J. Pennycook et al. showed how configuration differences affect the performance of a CUDA-based implementation of the Finite-Difference Time-Domain (FDTD) method~\cite{livesey2012development}.
The FDTD method provides a robust, straightforward way to solve Maxwell's curl equation. They considered differences in memory access patterns, single- and double-precision arithmetic, and hardware generation.
Their experimental results show that when the ratio between the number of threads per block and the number of available cores in the GPU is about 0.15, their method has significantly more efficient global memory access in T10. Furthermore, their method significantly benefits from the special handling of 64-bit accesses in the M2050 GPU, with only 1.3 times more double-precision computations than single-precision computations.

%With the further improvement of CUDA, the Nvidia GPUs also being updated rapidly. Newer researches were mostly done based on newer hardware to achieve better acceleration.
In 2016, Md. Maruf Hussain et al. implemented Standard Particle Swarm Optimization (SPSO) based on CUDA ~\cite{hussain2016cuda}. Experiments were conducted on an NVIDIA GeForce GTX 980 GPU and a single-core 3.20 GHz Intel Core i5 4570 CPU, and the test results show that the GPU algorithm runs up to 46 times faster than the corresponding CPU algorithm.

In 2017, Ahmed A. Abdelrahman et al. applied CUDA to the Advanced Encryption Standard (AES) in cryptography~\cite{abdelrahman2017high}. AES encryption and decryption are very time-consuming for a large amount of data. They implemented AES-128 ECB encryption on three GPU architectures (Kepler, Maxwell, and Pascal). Moreover, they also optimized the algorithm for some key factors, including crucial memory locations, computation granularity, and thread block size.
Their investigation showed that other researchers' implementations on Tesla, Fermi, and Kepler achieved speedups ranging from 47 Gbps to 68.6 Gbps~\cite{mei2010cuda,osvik2010fast,nishikawa2012high,li2012implementation,gilger2012gpu,nishikawa2014throughput}.
%\cite{nishikawa2012high}\cite{li2012implementation}\cite{gilger2012gpu}\cite{nishikawa2014throughput}. 
Their work achieved encryption speeds of 207 Gbps on NVIDIA GTX TITAN X (Maxwell) and 280 Gbps on NVIDIA GTX 1080 (Pascal) by implementing new optimization techniques using 32 bytes/thread granularity. 

Paulo A.C.Lopes et al. implemented the Hungarian algorithm(HA) using CUDA in 2019, which is one of the solutions to the linear assignment problem(LAP)~\cite{lopes2019fast}. LAP can be widely used in the traveling salesman problem~\cite{lawler1986erratum}, tracking in image processing~\cite{vasconcelos2009bipartite}, linear algebra~\cite{sathe2012auction}, telecommunications~\cite{yin2000efficient}, etc. They used several blocks of the GPU for alternating path-searching phases of the algorithm and parallelized the traversal of the graph. The test results on the GeForce GTX 970 GPU show that the parallel version can achieve up to 32 times faster than the serial version.
In 2020, Abir Al Sideiri et al. designed and implemented the Fisher classification scheme using CUDA in 2020~\cite{al2020cuda}. This is a time-consuming step in the image fractal compression algorithm. Encoding time, compression ratio, and peak signal-to-noise ratio are used as metrics to evaluate the correctness and performance of the developed algorithm. 
Test results on NVIDIA GeForce GT 660 M GPU show that using CUDA can achieve up to 6.4x speedup compared to serial algorithms.
Haythem Bahri et al. proposed a new implementation of a moving body detection algorithm on GPUs based on CUDA~\cite{bahri2020real}. Moving objects are first extracted by background subtraction based on GPU Gaussian Mixture Model (GMM). Then, two complementary features are extracted for moving object classification. The implementation of this algorithm on the GPU is 19 times faster in execution time than on the CPU.
Vincent Delmas et al. developed a multi-GPU version of a time-explicit finite volume solver for Shallow-Water Equations (SWE) using MPI in conjunction with CUDA-Fortran~\cite{delmas2020multi}. They adopted a CUDA-Aware OpenMPI version to speed up message passing between MPI processes. Experiments on two real domains with complex bathymetries show that comparing the multi-GPU version to the pure MPI multi-CPU version, about 100 CPU cores would be needed to achieve the same performance as one GPU.
In 2021, L. Antonelli implemented a modified version of the Smoothed Particle Hydrodynamics (SPH) method using CUDA~\cite{antonelli2021cuda}. The use of fast summation in a parallel computing scheme can speed up Taylor series expansion. GPU implementation can provide up to 90 times faster execution time compared to CPU.

%\subsubsection{Benchmarks for evaluating CUDA programming}
\subsubsection{Evaluations for CUDA programming}
Creating benchmarks is a practical choice to evaluate some CUDA programming techniques or exhibit performance challenges. Benchmarks can be used to help users understand the complexity of heterogeneous GPU-accelerator systems and guide users for performance optimization.

Rodinia is a benchmark suite proposed by Shuai Che et al. in 2009 for heterogeneous computing~\cite{che2009rodinia}. CUDA and OpenMP are used in Rodinia to explore multi-core CPUs and GPUs.
Rodinia is structured to span a range of parallelism and data sharing characteristics and can represent different types of behavior according to the Berkeley dwarves.
Its significance is to provide the benchmarks for testing the performance of multi-core CPUs and GPUs and get some essential architectural insights through the experimental results.

Using CUDA, M Abdullah Shahneous Bari et al. have evaluated a suite of benchmarks including Ray-tracing, Matrix-Matrix Multiplication, Heat Transfer, and sparse matrix-vector multiplication in different generations of NVIDIA GPUs, including Kepler, Maxwell, Pascal, and Volta~\cite{bari2018data} in 2018. The primary evaluation is for different GPUs, the performance difference of data placed in different memory including shared memory, constant memory, texture memory, etc.

%improvements

\subsubsection{Performance issues and the improvement}

Although CUDA can be applied to many areas to improve performance, programming to achieve high performance using CUDA has been challenging.
A GPU has hundreds or thousands of cores that a program must exhibit sufficient parallelism to achieve maximum GPU utilization.
Also, a GPU accelerator system has a heterogeneous and deep memory system that programmers must effectively and correctly use to take advantage of the GPU's parallelism capability fully.

The manycore architecture and the complexity of GPU heterogeneous memory architecture and system indicate three important guidelines for developing high-performance CUDA programs. 
\begin{itemize}
    \item First, GPU kernels should be optimized to saturate the massively parallel capability of GPUs as much as possible. 
    \item Second, the deep memory hierarchy inside GPU should be effectively leveraged to maximize the computing efficiency of GPU for kernel execution.
    \item Third, memory management and data movement between CPU and GPU memory should be properly arranged to reduce the performance impact of data movement operations.
\end{itemize}
%We will analyze the performance challenges and collect the related work of solving these challenges.

\paragraph{Using CUDA kernel optimization techniques to optimize the performance}

Based on the first guideline, there are several techniques to optimize the GPU kernels, including avoiding warp divergence, dynamic parallelism, concurrent kernel, and task graphs.

\begin{enumerate}
\item[-] \textit{Avoiding warp divergence}
\end{enumerate}

A warp is the scheduling unit of GPU kernel execution, made up of a group of threads within a block. Because of GPU's lock-step SIMT execution model, when two threads in a warp execute two separate branches during execution, there could be a significant performance penalty. This divergence of threads because of branching is known as warp divergence.
In the study of SNN simulations, Jayram Moorkanikara Nageswaran et al. found that warp divergence can occur if different threads within the same warp take different paths after a branch conditions~\cite{nageswaran2009efficient}. If the code executed after a diverging condition is simple, then the effect due to warp divergence is minimal.
Therefore, they reduce the effects of warp divergence by buffering the information used to diverge loop execution and delaying execution until there is enough data for all threads to execute.
Meng et al. propose compiler side optimizations
to dynamically handle warp divergence\cite{meng2010dynamic}.

\begin{enumerate}
\item[-] \textit{Using dynamic parallelism}
\end{enumerate}

The dynamic parallelism technique allows the GPU to generate its work by launching another kernel instead of requiring the kernel jobs submitted by the CPU.
The work proposed by J DiMarco et al. ~\cite{clustering} in 2013 to optimize K-means
clustering and hierarchical clustering using dynamic parallelism. 
In the same year, Jianqiang Dong et al. implemented BIRCH (one of the most famous streaming data clustering techniques) using dynamic parallelism~\cite{dong2013accelerating}. Their experimental results on NVIDIA Tesla K20 show that the GPU-accelerated BIRCH can be 154 times faster than the CPU version on multiple benchmark problems, with good scalability and high accuracy.
Their team also proposed a parallel method for traversing DFS code trees and developed a parallel gSpan with dynamic parallelism~\cite{wang2013graph}. Their method achieves up to 80x speedup for traversing DFS code trees, and parallel gSpan outperforms the original gSpan by the order of magnitude.

\begin{enumerate}
\item[-] \textit{Using concurrent kernel}
\end{enumerate}

From Fermi architecture, NVIDIA GPUs introduced a "concurrent kernels" feature, which enables the CPU to launch multiple kernel instances on one GPU.
In 2011, Masashi Oiso et al. implemented a steady-state genetic algorithm (GA) on GPU based on the concurrent kernel technique, which was 3.0-6.0 times faster than the corresponding CPU implementation~\cite{oiso2011accelerating}.
In 2012, Florian Wende et al. mentioned that the concurrent kernel technology has performance degradation when multiple host threads invoke multiple GPU kernels in succession without synchronizing their actions. They proposed a producer-consumer principal approach to manage GPU kernel calls within parallel host regions by reordering individual GPU kernels before actually calling them~\cite{wende2012improving}. They demonstrated their method's effectiveness in an intense scaling simulation of a small molecule solvated within a nanodroplet.
In 2020, S Shekofteh et al. studied how to implement concurrent kernel more efficiently, including a framework for scheduling\cite{ShekoftehConcurrent}.

\begin{enumerate}
\item[-] \textit{Using task graphs}
\end{enumerate}

The task graph feature is pretty new, and it was introduced in the 2018 release of CUDA 10. The task graph can consist of a series of operations, such as memory copy and kernel startup, connected through dependencies and defined separately from their execution.
This technique allows the CUDA runtime to perform whole-graph optimization and significantly reduce the kernel call overheads. 
In 2019, L Toldeo et al. who studied the performance impact of task graphs in scheduling,
focusing on leveraging concurrent kernel and dynamic parallelism as the kernels to be scheduled with a task graph\cite{ToledoTaskGraph}. 
The authors claim a 25x performance uplift when using task graphs, which illustrates how combining multiple features can significantly improve performance.
In 2021, Dian-Lun Lin et al. proposed a lightweight task graph programming framework to help developers avoid setting complex parameters and managing graphs with complex dependencies~\cite{lin2021efficient}. Their framework works well in microbenchmarks and large-scale machine learning workloads. Moreover, it also achieves a very similar performance to an optimally-constructed graph and consumes much fewer GPU resources.

\paragraph{Effectively Leveraging the Deep Memory Hierarchy Inside GPU to Maximize The Computing Capability of GPU for Kernel Execution}

According to the second guideline, effectively leveraging the deep memory hierarchy inside GPU is well-studied, including using shared memory, using coalesced memory access, memory alignment for the memory access, overlapping and pipelining data copy between global memory and shared memory using \texttt{memcpy\_async}, using data shuffle between thread and avoiding bank conflicts.

\begin{enumerate}
\item[-] \textit{Using shared memory }
\end{enumerate}
Shared memory is a high-speed programmable SRAM on the GPU chip, and all threads in the same block can access it. 
The latency of the shared memory is almost the same as that of the register, and its capacity is several times that of the register. Moving the data that must be repeatedly accessed to the shared memory can improve the program's performance.
Zhiyi Yang et al. used shared memory to improve the data reading speed of the DCT encode and decode algorithm in image processing\cite{yang2008parallel}. Paulius Micikevicius et al. optimizes the stencil with shared memory to significantly increase the speed of 3D finite-difference computation\cite{micikevicius20093d}.
The caching algorithm for CUDA shared memory for 2D Smoothed particle hydrodynamics (SPH) solver implementation proposed by Daniel Winkler et al. can significantly improve the efficiency of the original algorithm\cite{winkler2017gpusphase}.

\begin{enumerate}
\item[-] \textit{Using coalesced memory access}
\end{enumerate}

On a GPU, data transfer between global memory and on-chip storage is by chunk for each memory transaction, even though a thread requests only a small subset of a chunk.
Memory coalescing is a technique in GPU programming to use minimum transactions to fulfill the memory requests by a large number of threads.
This is accomplished by combining references to adjacent memory locations of multiple threads into a minimum number of transactions. 
In 2009, N. S. L. Phani Kumar et al. used CUDA to parallelize an iterative algorithm, Expectation Maximization (EM), which is widely used in the fields of signal processing and data mining~\cite{kumar2009fast}.
Their experimental results show that coalesced memory access can improve the kernel by 250x.
In 2011 Yuri Torres et al. discussed the performance impact of coalesced memory access on the Fermi architecture~\cite{torres2011understanding}. Using matrix addition and matrix multiplication, they showed how to implement coalesced memory access and tested the performance of both applications. They point out that using coalesced memory accesses can improve cache hit rates.
As mentioned in section~\ref{subsec:Implementationsindifferentareas}, Md. Maruf Hussain and others implemented SPSO based on CUDA~\cite{hussain2016cuda} in 2016, and they also used coalesced memory access to optimize performance further. Video RAM (VRAM) bandwidth is most efficiently used when coalesced memory access is used.
\begin{enumerate}
\item[-] \textit{Memory alignment for the memory access}
\end{enumerate}

Aligned memory access means that the first memory address accessed is the exact multiple of a memory chunk.
The test results in the modified AXPY example show that the aligned access has a clear though small performance improvement (about 3\%) because a smaller number of memory transactions are performed in the aligned memory access situation than in the misaligned memory access situation for the same amount of needed data~\cite{yi2021cudamicrobench}. 

\begin{enumerate}
\item[-] \textit{Overlapping and pipelining data copy between global memory and shared memory using \texttt{memcpy\_async}}
\end{enumerate}

Before using shared memory, data needs to be copied from slower global memory to the faster shared memory. 
To accelerate this copy, recent CUDA introduced asynchronous memcpy, the  {\it memcpy\_async} function~\cite{choquette2020nvidia},  for further optimize data movement between global memory and shared memory. This feature is realized based on two aspects in NVIDIA’s new Ampere Architecture, being the hardware acceleration of the {\it memcpy\_async} operation that bypasses register access and pipelining the {\it memcpy\_async} operation to overlap computation and memory operations.
The hardware acceleration of the memory operation is done by bypassing temporary registers when copying from global to shared memory. This hardware-level optimization provides a small but meaningful performance improvement.

\begin{enumerate}
\item[-] \textit{Using data shuffle between threads}
\end{enumerate}

Kepler and new generation GPUs introduce a new warp-level intrinsic called the shuffle operation. This feature allows threads in the same warp to exchange data directly between registers, bypassing the local memory. 
In 2015, Yongchao Liu et al. developed LightSpMV, a new CUDA-compatible SpMV algorithm using the standard CSR format~\cite{liu2015lightspmv}. LightSpMV, with atomic operations and warp shuffle functions as fundamental building blocks, so two dynamic row distribution methods are investigated at vector and warp levels. On Tesla K40c GPU, LightSpMV outperforms CUSP and cuSPARSE.
In 2016, Wai-Kong Lee et al. presented the implementation of block ciphers in NVIDIA GTX 980 with Maxwell architecture~\cite{lee2016fast}.
They used a high-speed register to store the encryption keys and used shuffle to swap them between threads in the same warp using the warp shuffle operation to improve performance further.

\begin{enumerate}
\item[-] \textit{Avoiding bank conflicts}
\end{enumerate}

GPU shared memory is architectured into multiple equal-sized memory modules (banks). When shared memory is allocated, consecutive data are sequentially mapped to 32 consecutive banks in cyclic distribution.
When different threads in a warp access different locations of the same bank simultaneously, access is serialized. This is known as bank conflict, significantly impacting GPU kernel performance.
Y Kim et al. proposed CuMAPz to compare the memory performance of CUDA programs\cite{kim2011cumapz}. For some effects, including data reuse, global memory access coalescing, shared memory bank conflict in shared memory, they explained how CuMAPz optimizes memory strategy.

\paragraph{Properly Arranging Data Movement Between CPU and GPU to Reduce the Performance Impact of Data Movement for the GPU Program}

The third guideline is to arrange data movement between CPU and GPU properly. Commonly used optimization techniques are overlapping and pipelining GPU data copy and kernel computation using CUDA stream and \texttt{cudaMemcpyAsync}, storing read-only data in read-only memory, using unified memory for applications with low access density and reducing unnecessary data transfer.

\begin{enumerate}
\item[-] \textit{Overlapping and pipelining GPU data copy and kernel computation using CUDA stream and \texttt{cudaMemcpyAsync}}
\end{enumerate}

Data movement between CPU and GPU often dominates the total time of offloading a computation kernel to a GPU. A well-known solution is to use CUDA stream and {\it cudaMemcpyAsync} to move data between CPU and GPUs~\cite{guide2013cuda}. This approach enables parallelism and overlapping between data movement and kernel computation, thus would be ably decreasing the impact of data movement on the overall offloading performance.
In 2009, Paulius Micikevicius et al. used {\it cudaMemcpyAsync} in their work to accelerate 3D Finite Difference Computation\cite{micikevicius20093d}. For smaller data sets, communication overhead is close to or even greater than the computation time. Using {\it cudaMemcpyAsync} can significantly improve the efficiency of program execution.
In 2012, Juan G´omez-Luna et al. evaluated CUDA streams and cudaMemcpyAsync and pointed out that using streams in applications where input data instances are independent can be very profitable~\cite{gomez2012performance}.
In 2015, Mohammed Sourouri et al. developed a program for stencil computations based on CPU and GPU\cite{sourouri2015cpu+}. They used {\it cudaMemcpyAsync} to perform the final CPU-GPU data exchange to achieve a good overlap of various computing activities.

\begin{enumerate}
\item[-] \textit{Storing read-only data in read-only memory}
\end{enumerate}

NVIDIA GPUs have reserved part of the DRAM as read-only memory, known as constant memory and texture memory. 
Properly use read-only memory and global memory by storing read-only data in 
constant or texture memory and read-write data in global memory improve the usage of the bandwidth of DRAM. 
In 2009, D.P. Playne applied texture memory to three-dimensional Cahn-Hilliard simulations\cite{playne2009data}. The unique design of texture memory poses a significant advantage in processing three-dimensional data.
In 2011, Lucas C. G. G. Persoon et al. used texture memory to significantly reduce the calculation time of $\gamma$ evaluations without reducing the accuracy\cite{persoon2011fast}. 
This is a successful texture memory application in quantifying differences in dose distributions. 

\begin{enumerate}
\item[-] \textit{Using unified memory for applications with low access density}
\end{enumerate}

Memory access density refers to the ratio of sizes of the data used for calculation and the data transferred between CPU and GPUs.
When the access density is low, moving useless data will impact the performance. One of the optimized solutions is to use unified memory such that data are copied from CPU memory to GPU memory when needed and accessed. GPU unified memory transfers only the necessary pages which contain the necessary data between CPU and GPU during execution.
In 2014, R Landaverde et al. developed multiple microbenchmarks for the GPU architecture and tested the performance of Unified Memory Access(UMA)\cite{landaverde2014investigation}. 
In 2015, W Li\cite{li2015evaluation} used the Diffusion3D benchmark in the CUDA SDK samples, Parboil benchmark suite, and matrix multiplication to evaluate unified memory technology\cite{coombes2014tegra}.

\begin{enumerate}
\item[-] \textit{Reducing unnecessary data transfer}
\end{enumerate}
A simple and classic example of unnecessary data transfer is sparse matrix processing.
It is obviously not cost-effective to transfer all 0-elements in the sparse matrix to the GPU's memory.
Therefore, using compression storage formats such as Dictionary of keys (DOK), List of lists (LIL), Coordinate list (COO), Compressed sparse row (CSR) and Compressed sparse column (CSC) can greatly reduce the amount of data to be transferred and the amount of computation.

\subsection{Comparison of different programming methods on GPU and translations between them}
Exploiting the available performance of heterogeneous architectures may be challenging. As mentioned above, there are various parallel programming frameworks, including OpenMP, OpenACC and CUDA. Choosing the option appropriate to the target context is not straightforward.
Many researchers have made comprehensive evaluations of different methods in terms of performance, programming difficulty and portability to give suggestions to programmers.

The Standard Performance Evaluation Corporation(SPEC), is a company specializing in benchmarks. It includes a SPEC ACCEL, a set of compute-intensive parallel applications running under the OpenCL 1.1, OpenACC 1.0, and OpenMP 4.5 APIs. This suite comprehensively evaluates CPU and GPU performance, memory performance, and even compilers' performance.

In 2013, Tetsuya Hoshino et al. compared the performance of CUDA and OpenACC using two kernel benchmarks, matrix multiplication and 3-D stencil, and a real-world computational fluid dynamics (CFD) application~\cite{hoshino2013cuda}. Overall, the performance of using OpenACC is about 50\% that of using CUDA, up to 98\% depending on the compiler. For CFD applications, the performance of using OpenACC is about 36\% of using CUDA. This gap will be more evident if the CUDA program is deeply optimized. The reason was that the OpenACC specification at that time could not apply shared memory due to programming interface limitations.
In 2020, Jan Eichstädt et al. evaluate the Kokkos~\cite{calore2016performance}, OpenMP and OpenACC programming models in the context of higher-order methods for CFD~\cite{eichstadt2020comparison}. Aspects to compare are implementation effort and code maintainability and performance and portability across architectures. In terms of performance, the OpenACC model showed the best overall performance, yielding the fastest runtimes on Nvidia GPUs and only slightly slower runtimes than OpenMP on Intel CPUs.
In 2021, Mikhail Khalilov et al. compared the performance of CUDA, OpenMP, and OpenACC on state-of-the-art Nvidia Tesla V100 GPUs in some typical scientific programming scenarios and evaluated the performance of physical simulation code implemented using this programming models~\cite {khalilov2021performance}.
The test results show that for Copy, Mul, Add, Triad tests, using CUDA can achieve the maximum bandwidth, OpenACC and OpenMP on the same tests show the same lag behind CUDA with an average of 3-4\% and 6-7\%, respectively.
On the Dot test, using CUDA can reach 91\% of the peak bandwidth. Using OpenMP can reach 14\% of the peak bandwidth and using OpenACC can reach 76\% of the peak bandwidth. 
Their conclusions are very representative. OpenMP and OpenACC compilers can generate efficient parallel code for simple cases, and as code complexity increases, CUDA shows significant performance benefits. A similar trend was observed while benchmarking Cloverleaf and LULESH applications.

%Put Translation here:

The advantage of OpenMP target offload compared to CUDA is that it can easily offload the computation to the device.
In 2020, Christopher Daley et al. discussed how to convert CUDA programs in benchmark HPGMG to OpenMP target offload and try to use explicit data management~\cite{daley2020case}. 
Their experimental results showed that the XL compiled managed memory version of HPGMG achieves 0.70x the performance of HPGMGCUDA on Summit, and the LLVM/Clang compiled explicit data management version HPGMG achieves 2.04x of HPGMG-CUDA on Cori-GPU and 0.73x of HPGMG-CUDA performance on Summit.

%\section{Comparasion of different parallel programming}

%The Standard Performance Evaluation Corporation(SPEC), is a company specializing in benchmarks. It includes a SPEC ACCEL，which has a set of compute-intensive parallel applications running under the OpenCL 1.1, OpenACC 1.0, and OpenMP 4.5 APIs. This suite comprehensively evaluates CPU and GPU performance, memory performance, and even compilers' performance.

\section{Single Instruction Multiple Data} 
\label{sec:simd}
SIMD (single-instruction multiple-data) or vector architecture has been an effective parallel processing solution to address the plateauing of Moore’s law and the need for high performance of scientific data processing.
SIMD uses long registers with special instructions that can perform several instances of a certain operation at once in the same amount of CPU time as a single operation. Often referred to as vector processing, such an approach is most advantageous for parallelizing loops, which often consumes the most time in many computation-intensive applications.
The majority of programming languages, including the most common- C, C++, and Fortran- do not have any constructs that map directly to vector programming. As a result, either specialized functions are needed, or the compilers have to find a way to do vectorization themselves.
We still discuss vectorization in three categories: auto-vectorization by compilers, explicit vectorization such as OpenMP, and programming with vector intrinsics. 

\subsection{Auto-Vectorization}
Auto-vectorization is the ideal method and probably the most researched one. Optimizing compilers such as GCC, LLVM, and other vendor compilers do a decent job of analyzing and vectorizing simple data-parallel code. However, they have challenges of vectorizing loops that require information that is hard to analyze by the compiler. For example, a data-parallel loop with a reduction operation has loop-carried dependencies. 
While the mainstream compilers may do a degree of vectorization on their own, they typically require some sort of user intervention, such as architecture flags to auto-vectorize to the greatest extent. From our own experience, we have found that the compilers often do not optimize fully to the hardware available without several particular flags- for example, Clang often generated only AVX-2 or even SSE4 instructions despite AVX-512 hardware being available.

In 2006, Dorit Nuzman et al. developed a compiler containing a new generic vectorization technique for interleaved data, which can effectively vectorize non-contiguous access patterns with a constant stride of power of 2~\cite{nuzman2006auto}. Today's SIMD models do not directly support operations on disjoint vector elements. Once interleaved data is correctly reorganized, it will significantly benefit from SIMD. Their experimental results show that for interleaving levels (stride) as high as 8, the speedups in execution time can be up to 3.7.
They also developed an automatic vectorization program in GCC~\cite{nuzman2006multi}. Their design can be adapted to various SIMD architectures, and different alignment mechanisms are designed for different SIMD platforms. Experimental data conducted on four different SIMD platforms showed that their design achieved significant speedups on key cores.

In 2011, Olaf Krzikalla et al. proposed a vectorization tool named Scout.
Scout is designed as a source-to-source translator that generates code with SIMD intrinsics~\cite{krzikalla2011scout}. They implement the vectorization optimization technique based on the syntax tree generated by the Clang parser. The unroll-and-jam technique is applied to vectorize the loop body, supporting multiple SIMD instruction sets, including SSE and AVX2. In the experiments, they observed that the overall speedup was nearly 1.5.

In 2017, Oliver Reiche et al. proposed automatic vectorization technologies for image processing DSL in the context of source-to-source compilation and integrated these technologies into the open-source DSL framework~\cite{reiche2017auto}. Compared with the non-vectorized execution using the latest C/C++ compiler, the geometric average speed of benchmarks obtained from ISPC and image processing has increased by 3.14.
In the same year, Matthew Lambert et al. demonstrated how the auto-vectorization capabilities of Clang/LLVM and GCC could be used to enhance the performance of the Four Russians Matrix Multiplication problem~\cite{lambert2017compiler}. They used a method to store multiple small prime numbers into a single word called bit-packing. This, in combination with vectorization, yielded significant speedups.

In 2020, Ameer Haj-Ali et al. proposed an end-to-end method that can vectorize loops using deep reinforcement learning(RL)~\cite{haj2020neurovectorizer}. They integrated RL in the LLVM compiler and used deep learning to dynamically capture different instructions, dependencies, and data structures to determine the optimal vectorization factor. Their experiments show that the performance can be increased up to 4.73 times compared with the baseline.

With the development of vectorization technology, many researchers are devoted to solving some performance bottlenecks. 

In 1994, Ken Kennedy et al. proposed a context optimization method for SIMD execution~\cite{kennedy1994context}.
Disabling the processor that is not involved in the current calculation by changing the machine context is a problem that the SIMD compiler must solve. They optimize context partitioning and context splitting to reduce the cost of context changes. Their method can reduce the execution time of the hand-optimized MPL version by 12\%.

It is considered not easy for automatic vectorization tools to obtain good performance in the outer loop.
Dorit Nuzman et al. demonstrated vectorization for outer loops~\cite{nuzman2008outer}. They re-studied the method of outer loop vectorization and paid attention to the properties of modern short SIMD architecture. Compared with the innermost loop vectorization, which can only provide an acceleration factor of 1.53, the outer loop vectorization can significantly improve performance with a factor of 3.13x.

Angela Pohl et al. addressed potential performance concerns for the vector length agnostic (VLA) model introduced by Arm and added to RISC-V~\cite{9188238}. They showed that VLA code reaches 90\% that of fixed length models (i.e., Intel). Additionally, they showed that performance does not increase proportionally with higher vector lengths due to higher memory demands.

\subsection{OpenMP SIMD directive}

We consider explicit vectorization an ideal solution to auto-vectorization, arguably the middle of auto-vectorization and manual vectorization. It gives the end-user programmer control and transparency over the code without forcing the compiler to do anything it cannot.
While some compilers such as Clang/LLVM and the Intel compiler provide directives that can perform this explicit vectorization, solutions such as OpenMP's SIMD instructions provide a standard, cross-platform cross-compiler solution. OpenMP SIMD directives are used for loops to specify code that should be vectorized. The user can fine-tune vectorization using various clauses, specifying vector lane widths, safe widths, alignment, and more.

Michael Klemm first proposed the OpenMP SIMD instruction in 2012~\cite{klemm2012extending}. The SIMD instruction will allow programmers to guide the vectorization process, thereby achieving more efficient and portable use of the SIMD level.
In 2016, Florian Wende et al. investigated the ability of current compilers (GNU, Clang, and Intel) to generate SIMD code for some common microbenchmarks and two cores for VASP and MOM5/ERGOM applications~\cite{wende2016portable}. They also explored coding strategies to improve SIMD performance across different compilers and platforms. They concluded that OpenMP 4.x SIMD instructions can achieve portable performance in many cases.
They proposed a portable SIMD coding scheme called "enhanced
explicit vectorization" for which different compilers can understand and generate code equally. 
The microbenchmarks they developed show that on Haswell, SIMD execution gives speedups between 2x and 4x.
In 2017, Jinpil Lee et al. proposed a programming interface that connects user-defined SIMD functions and SIMD loops~\cite{lee2017extending}.
Also, they introduced a new directive called alias simd that specifies a user-defined SIMD variant of the target function. Using this interface, the loop iteration translation and the SIMD code generation of the loop body are split. Users can write highly optimized SIMD code using architecture-dependent programming methods such as intrinsics functions.
Although the primary target architecture for this interface is ARM SVE, they also considered AVX.
In 2021, Joseph Huber et al. introduced a methodology that uses LLVM-based tools that aim to tune DCA++ (Dynamic Cluster Approximation) applications to the new ARM A64FX processor~\cite{huber2021case}. The goal is to describe the changes required for the new architecture and generate efficient single-instruction/multiple-data (SIMD) instructions for the new SVE instruction set. They refactored the code to apply SIMD-optimized transformations and ensured the correctness and optimal performance of the transformations.
By applying these changes, the code is 1.98 times faster.

\subsection{Programming with Vector Intrinsics}

Vector intrinsics are compiler-provided pseudo-functions that map 1:1 to the underlying assembly or are very close to that.
For experienced programmers, it is possible to get higher optimizations than an optimizing compiler. Moreover, this way can guarantee the conversion.
However, its disadvantage is also obvious. The vectorization by intrinsics is not cross-platform. If we write a vectorized loop using AVX-512 intrinsics, it will not work for Arm SVE. This results in its poor portability. Second, good vectorization requires time and skill, which is not something programmers always have or should need to have.

In order to facilitate developers to obtain high performance from SIMD, in 2014 quickly, Haichuan Wang et al. developed a general-purpose SIMD library, an open-source, portable C++ interface~\cite{wang2014simple}. It supports most C++ operators and provides various mappings from platform-specific intrinsics to generic SIMD intrinsics. Code developed based on this library is portable across different SIMD platforms. Compared to platform-specific intrinsic code, their library achieved similar performance with 22\% less code on average.
In 2018, Wenge Liu et al. compared the performance of using coarse-grained OpenMP vectorization schemes and the 256-bit AVX instruction set for the memory/computation-intensive acoustic wave equation with CPU template buffer optimization~\cite{liu2019parallel}. They applied 8-channel parallel vectors to simulate seismic wavefields to increase the computational bandwidth, thereby eliminating many computational instructions and achieving a speedup of 3-7 times.
In 2019, Enzo Rucci et al. proposed a faster way to compute the Smith-Waterman (SW) algorithm~\cite{rucci2019swimm}.
The first optimized the 32-bit integer version to improve instruction-level parallelism utilization on the KNL architecture and enabled 64 SIMD channel utilization via the AVX-512BW ISA. Their experimental results show that the AVX2 instruction set remains the best choice for Intel's current HPC platforms. The peak performances on KNL accelerators Xeon Gold 6128 and Xeon Gold 6138 are 511, 306.3 and 734 GCUPS, respectively.

\section{Conclusion} 
\label{sec:conclusion}

In this survey, we investigate three commonly used ways to achieve parallel computing on different platforms, including taking advantage of the CPU multithreading technique on multi-core CPUs, using GPU-accelerated heterogeneous systems, and using SIMD as an effective parallel processing solution.
For each way, we discussed three categories of programming techniques based on the programming complexity.
The first level is to use the compiler's automatic parallelization or directly use pre-defined libraries. The second level is explicit parallelism, which uses a directive-based programming language to parallelize certain parts of a program, like OpenMP. The third level is manual parallelization, which is using some special programming models or intrinsics to perform more direct operations on the hardware.
For all programming techniques, the applications in different areas, performance bottlenecks, solutions, and programming difficulties are investigated.

The survey can help in the following two aspects:
\begin{itemize}
   \item We can get some information on performance analysis, applicable scenarios, comparison, and evaluation of different parallel schemes.
   \item We can summarize the advantages and disadvantages of different methods. Then we can help programmers measure the trade-off between different methods, avoid common problems and provide optimization suggestions. For example, for CUDA, AVX512, etc., algorithms and programming skills significantly impact performance. For directive-based languages such as OpenMP and OpenACC, the impact of the compiler cannot be ignored. 
\end{itemize}

\bibliographystyle{IEEEtran}
\bibliography{references}

\end{document}